\documentclass[usenatbib]{mn2e}
\usepackage{graphicx} 
\usepackage{times} 

\title[A possible black hole in the gamma-ray microquasar LS~5039]
{A possible black hole in the gamma-ray microquasar LS~5039}

\author[J. Casares et al.]
{J. Casares$^1$\thanks{E-mail: jcv@iac.es (JC); mribo@discovery.saclay.cea.fr (MR); iribas@ieec.uab.es (IR); jmparedes@ub.edu (JMP); jmarti@ujaen.es (JM); ahd@iac.es (AH)},
M. Rib\'o$^{2,3\star}$, I. Ribas$^{4,5\star}$, J.M. Paredes$^{6\star}$, 
J. Mart\'\i{}$^{7\star}$, A. Herrero$^{1,8\star}$\\
$^1$ Instituto de Astrof\'{\i}sica de Canarias, 38200 La Laguna, Tenerife, 
Spain\\
$^2$ DSM/DAPNIA/Service d'Astrophysique, CEA/Saclay, B\^at. 709, L'Orme des
Merisiers, 91191 Gif-sur-Yvette, Cedex, France\\
$^3$ AIM - Unit\'e Mixte de Recherche CEA - CNRS - Universit\'e Paris VII
- UMR n$^{\rm o}$ 7158\\
$^4$ Institut de Ci\`encies de l'Espai -- CSIC, Campus UAB, Facultat de Ci\`encies, Torre C5 - parell - 2a planta, 08193 Bellaterra, Spain\\
$^5$ Institut d'Estudis Espacials de Catalunya (IEEC), Edif.
Nexus, C/Gran Capit\`a, 2-4, 08034 Barcelona, Spain\\
$^6$ Departament d'Astronomia i Meteorologia, Universitat de Barcelona, Av.
Diagonal 647, 08028 Barcelona, Spain\\
$^7$ Departamento de F\'{\i}sica, Escuela Polit\'ecnica Superior, Universidad 
de Ja\'en, Campus Las Lagunillas s/n, Edif. A3, 23071 Ja\'en, Spain\\
$^8$ Departamento de Astrof\'{\i}sica, Universidad de La Laguna, Avda.
Astrof\'{\i}sico Francisco S\'anchez s/n, 38271 La Laguna, Tenerife, Spain}

\begin{document}

\maketitle

\begin{abstract}
The population of high energy and very high energy gamma-ray sources, detected
with EGRET and the new generation of ground-based Cherenkov telescopes,
conforms a reduced but physically important sample. Most of these sources are
extragalactic (e.g., blazars), while among the galactic ones there are pulsars
and supernova remnants. The microquasar LS~5039, previously proposed to be
associated with an EGRET source by Paredes et al. (2000), has recently been 
detected at TeV energies, confirming that microquasars should be regarded as a
class of high energy gamma-ray sources. To model and understand how the
energetic photons are produced and escape from LS~5039 it is crucial to unveil
the nature of the compact object, which remains unknown. Here we present new
intermediate-dispersion spectroscopy of this source which, combined with
values reported in the literature, provides an orbital period of $P_{\rm
orb}=3.90603\pm0.00017$~d, a mass function $f(M)=0.0053 \pm 0.0009$~M$_\odot$,
and an eccentricity $e=0.35\pm0.04$. Atmosphere model fitting to the spectrum
of the optical companion, together with our new distance estimate of
$d=2.5\pm0.1$~kpc, yields $R_{\rm O}=9.3^{+0.7}_{-0.6}$~R$_\odot$, $\log
(L_{\rm O}/$L$_\odot)=5.26\pm0.06$, and $M_{\rm
O}=22.9^{+3.4}_{-2.9}$~M$_\odot$. These, combined with our dynamical solution
and the assumption of pseudo-synchronization, yield an inclination
$i=24.9\pm2.8 \degr$ and a compact object mass $M_{\rm
X}=3.7^{+1.3}_{-1.0}$~M$_{\odot}$. This is above neutron star masses for most
of the standard equations of state and, therefore, we propose that the compact
object in LS~5039 is a black hole. We finally discuss about the implications
of our orbital solution and new parameters of the binary system on the CNO
products, the accretion/ejection energetic balance, the supernova explosion
scenario, and the behaviour of the very high energy gamma-ray emission with
the new orbital period.
\end{abstract}


\section{Introduction} \label{introduction}

The third EGRET catalog contains 271 sources detected at energies above
100~MeV \citep{hartman99}. Apart from extragalactic sources at high galactic
latitudes and some galactic pulsars and supernova remnants, the majority of
these sources, $\sim$168 or $\sim$62 per cent, still remains unidentified.
Among them, there are 72 sources located at low galactic latitudes, having
$|b|$$<$10$\degr$, which represents around 45 per cent of the unidentified
sources. Therefore, several of these objects are presumably of galactic
nature. The existence of similar properties between some of these sources,
indicate that there are at least three different groups of galactic
populations: a group of young stellar objects and star-forming regions
\citep{romero01}, the sources forming a halo around the galactic center, and
finally a group of sources correlated with the Gould Belt \citep{grenier00}
(see \citealt{romero04} and \citealt{grenier04} for recent updates). The
identification of their nature is of prime importance in high energy
astrophysics (see \citealt{cheng05} for a recent compilation).

On the other hand, microquasars are galactic X-ray binaries with relativistic
collimated jet emission. The compact object (a neutron star or a black hole)
accretes matter from a companion star, which can be a massive early type star
or a low-mass late type star. Since the first discoveries, one decade ago,
microquasars have raised a strong astrophysical interest because they provide
small scale counterparts of the highly energetic outflows seen in AGNs and
quasars. Their short characteristic time-scales allow us to study the physics
of accretion and outflow in a much faster way than in AGNs. Furthermore,
because of their binary nature, one can derive accurate orbital parameters and
masses for the accreting compact object and its donor star, which may be
correlated with jet properties. There are currently around 15 microquasars
known (\citealt{ribo05}) from a population of $\sim$300 X-ray binaries
(\citealt*{liu00}, \citeyear{liu01}), but only a few with well determined
system parameters. Detailed reviews on microquasars can be found in
\cite{mirabel99} and \cite{fender05}.

LS~5039 is the optical counterpart of the X-ray source RX~J1826.2$-$1450 and
it was proposed as a High Mass X-ray Binary (HMXB) by \cite{motch97}. Based on
archival NVSS data and their own VLA observations \cite*{marti98} identified
its radio counterpart, which appeared to be a persistent non-thermal radio
source. Subsequent VLBA observations, performed by \cite{paredes00}, revealed
relativistic radio jets at milliarcsecond scales, which qualified LS~5039 as a
microquasar. In the same work, Paredes et al. draw a possible connection with 
the unidentified gamma-ray source 3EG~J1824$-$1514, and suggest, for the first
time, that microquasars could be sources of high-energy $\gamma$-ray emission
above 100~MeV (see \citealt{paredes05} and \citealt*{riboetal05} for reviews
on other associations). Moreover, LS~5039 has been associated very recently
with the very high energy gamma-ray source HESS~J1826$-$148, detected at
energies above 250~GeV, reinforcing the association with the EGRET source
\citep{aharonian05}. The optical component of the LS~5039 system has been
classified as a O6.5\,V((f)) by \cite{clark01} and McSwain et~al.
(\citeyear{mcswain01}, \citeyear{mcswain04}; hereafter M01 and M04,
respectively) derived the first orbital parameters, with a proposed orbital
period of 4.4267 days and a highly eccentric orbit of $e=0.48\pm0.06$. 
In addition, evidence for intrinsic polarization at the 3 percent level was 
presented in \cite{combi04} and interpreted as Thomson scattering in 
the stellar envelope. 
The present paper reports novel spectroscopic observations of LS~5039 that 
support a revised orbital period of 3.9060 days, a higher mass function and
significantly lower eccentricity than previous claims. Our results have
important implications for the nature of the compact object, the presence of
CNO products in the optical companion, the mass loss of the supernova (SN)
explosion, the recoil velocity and runaway nature of LS~5039 (see
\citealt{ribo02} and M04), as well as for the interpretation of the existing
X-ray data (see \citealt{reig03}, \citealt{bosch05}, and references therein)
and very high energy gamma-ray data \citep{aharonian05}.

The observational details are outlined in section \ref{observations}. Section
\ref{radial} presents the analysis of the radial velocities and orbital
parameters while section \ref{rotational} focuses on the rotational velocity
calculation of the optical star. The determination of the stellar parameters
and chemical abundances are described in section \ref{stellar}. In section
\ref{mass} we present the implications of our new orbital solution for the
nature of the compact object. Based on all the updated parameters we conduct a
general discussion in section \ref{discussion}.

\section{Observations and Data Reduction} \label{observations}

We observed LS~5039 using the Intermediate Dispersion Spectrograph (IDS)
attached to the 2.5-m Isaac Newton Telescope (INT) at the Observatorio del
Roque de Los Muchachos on the nights of 23--31 July 2002 and 1--10 July 2003.
A total of 196 spectra were obtained with the combination of the 235~mm camera
and the R900V grating which provided a useful wavelength coverage (free from
vignetting) of $\lambda\lambda$3900--5500. We used integration times of
300--900~s. The seeing was variable (1--2 arcsec) during our runs and we used
a 1.2-arcsec slit which resulted in a resolution of 83~km~s$^{-1}$ (FWHM). In
addition, we also obtained five spectra of LS~5039 with the holographic
grating H2400B with the aim of measuring the rotational broadening ($v \sin
i$) of the optical companion's absorption lines. The spectral resolution of
these spectra was 30~km~s$^{-1}$ and we varied the central wavelength in order
to fully cover the range $\lambda\lambda$3900--5100, where most of the
prominent \mbox{He\,{\sc i}} and \mbox{He\,{\sc ii}} lines lie. The standard
star HD~168075~B, of spectral type O7\,V((f)), was also observed with the same
instrumental configurations for the purpose of rotational broadening analysis.
Furthermore, we obtained 14 H$\alpha$ spectra at 58~km~s$^{-1}$ resolution on
the nights of 4--7 July 2003 to derive the mass-loss rate in the optical star.
These spectra were also taken simultaneously with {\it RXTE} observations with
the aim of studying the correlation between X-ray flux and mass-loss rate
proposed by \cite{reig03} and confirmed by M04. These results are discussed in
\cite{bosch05}. A full observing log is presented in Table~\ref{log}.

\begin{table}
\centering
\caption[]{Log of the observations.}
\begin{tabular}{lcccc}
\hline 
Date          & Object      & No.  & Wav. Range & Disp. \\
              &            & spect.& $\lambda\lambda$ & (\AA/pix) \\
\hline
23--26 Jul 02 & LS~5039     &   4  & 4045--4910 & 0.85 \\
27--30 Jul 02 & "           &  67  & 3806--5865 & 0.63 \\
30 Jul 02     & HD~168075~B &   1  & "          & "    \\
31 Jul 02     & LS~5039     &  24  & "          & "    \\
~~~~~~"       & "           &   3  & 3920--4666 & 0.23 \\
~~~~~~"       & HD~168075~B &   1  & "          & "    \\
~~~~~~"       & LS~5039     &   2  & 4497--5224 & 0.22 \\
~~~~~~"       & HD~168075~B &   1  & "          & "    \\
1--10 Jul 03  & LS~5039     & 101  & 3806--5865 & 0.63 \\
4--7 Jul 03   & "           &  14  & 5350--6860 & 0.63 \\
\hline
\label{log}
\end{tabular}
\end{table}

The images were de-biased and flat-field corrected, and the spectra
subsequently extracted using conventional optimal extraction techniques in
order to maximize the signal-to-noise ratio of the output (\citealt{horne86}).
CuAr and CuNe comparison lamp images were obtained every 15--30 minutes, and
the $\lambda$-pixel scale was derived through 4/6th-order polynomial fits to
53/86 lines (depending on the set-up), resulting in an rms scatter
$<0.04$~\AA. The calibration curves were interpolated linearly in time.

\section{Analysis} \label{analysis}

\subsection{Radial Velocities and Period Search} \label{radial}

\begin{figure}
\centering
\includegraphics[angle=-90,width=84mm]{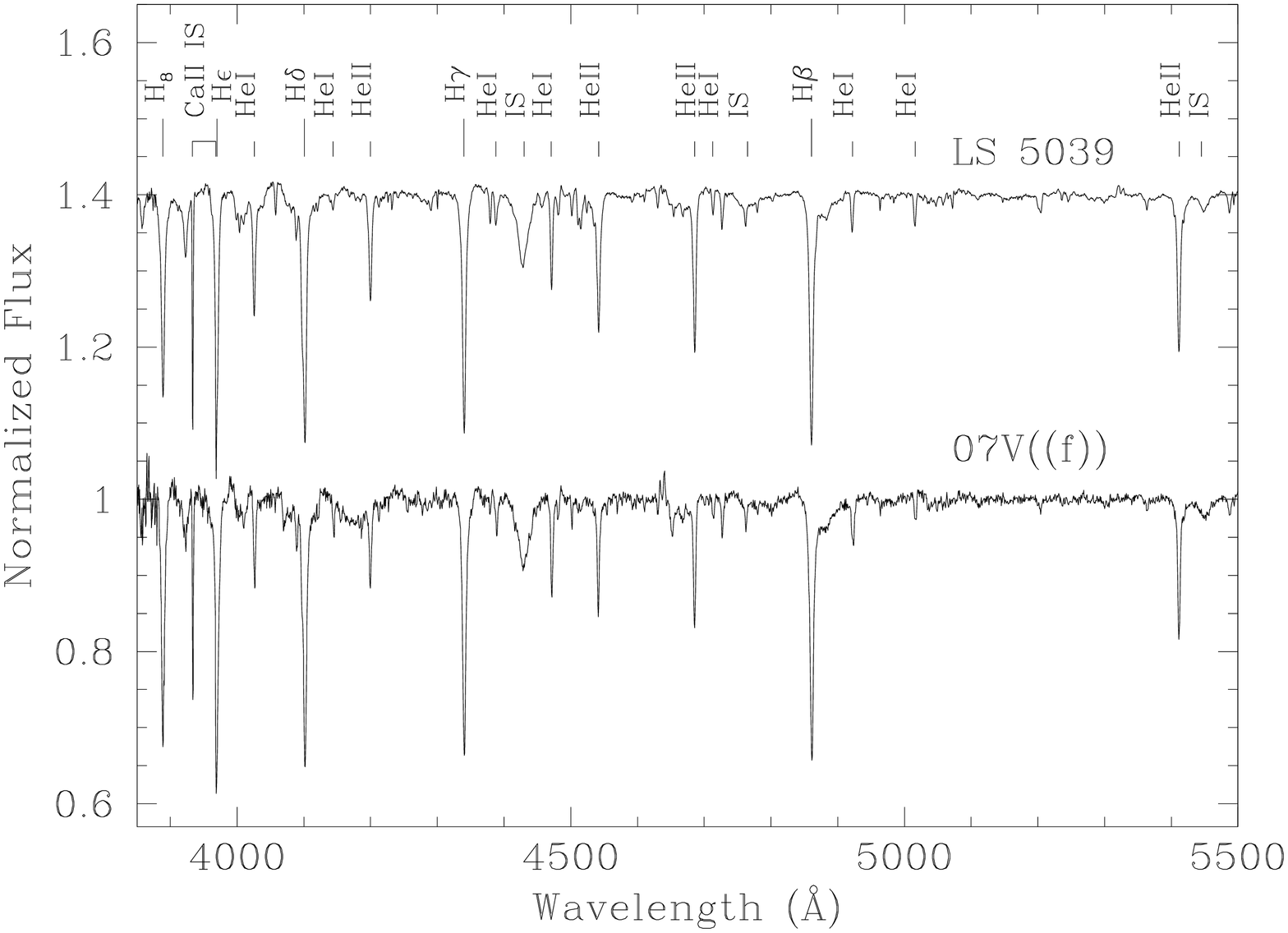}
\caption{Normalized average spectrum of LS~5039 (top, and offset 0.4 for 
display purposes) and the O7\,V((f)) star HD~168075~B (bottom). Main Balmer and
He lines are indicated. Note the steeper \mbox{He\,{\sc
i}}\,$\lambda$4471/\mbox{He\,{\sc ii}}\,$\lambda$4541 ratio for LS~5039, indicative of a slightly earlier spectral type than O7\,V.}
\label{figsp}
\end{figure}

\begin{figure}
\centering
\includegraphics[angle=-90,width=84mm]{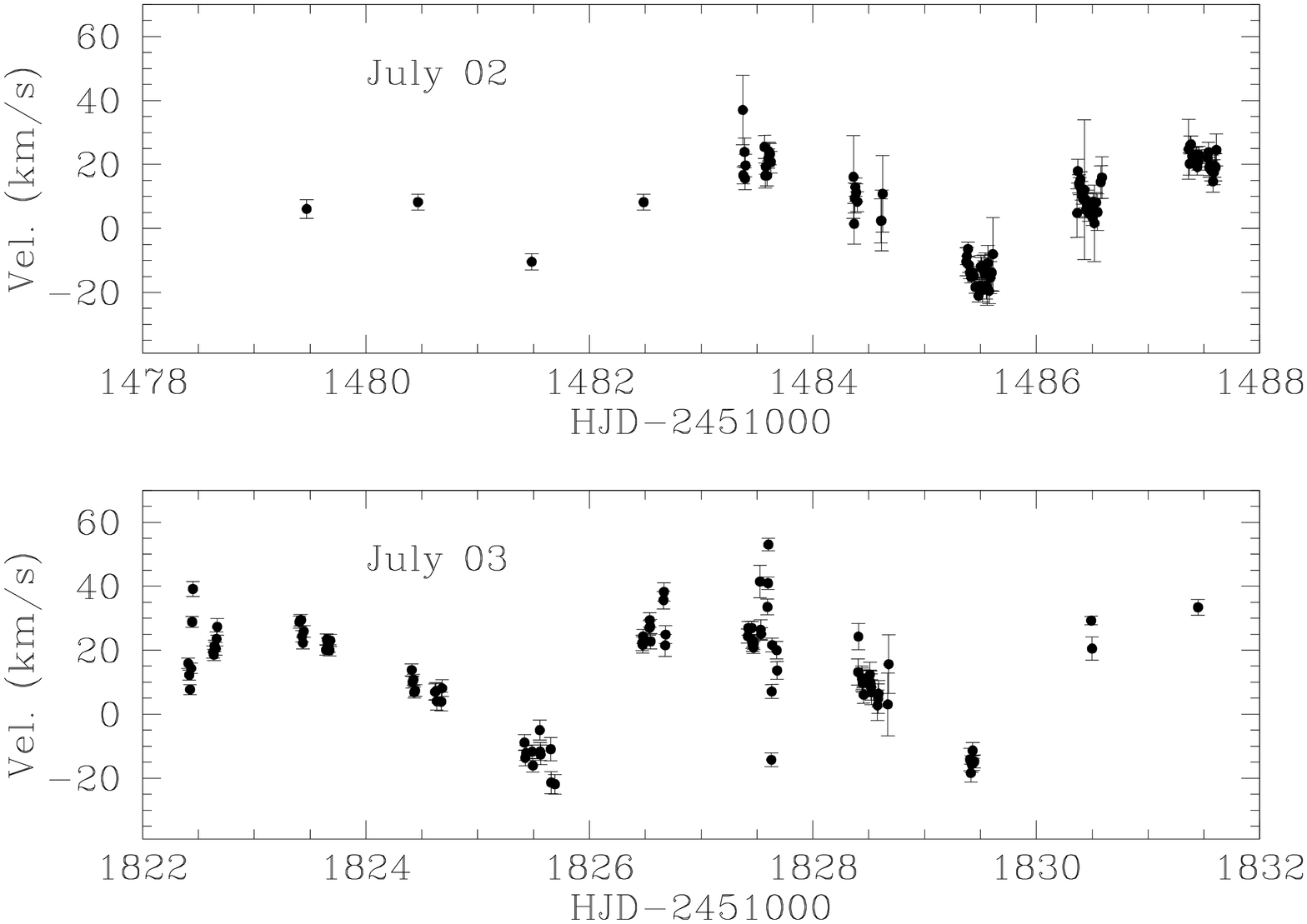}
\caption{Radial velocities of the 196 R900V spectra of LS~5039 obtained when
all the Balmer and He lines are included in the cross-correlation. A 4-day
modulation is clearly depicted.}
\label{figrvn}
\end{figure}

All the spectra were rectified, by fitting a low-order spline to the
continuum, and re-binned into a uniform velocity scale of 83~km~s$^{-1}$. We
show in Fig.~\ref{figsp} the averaged spectrum of LS~5039 along with the
template star HD~168075~B of spectral type O7\,V((f)). The ratio
\mbox{He\,{\sc i}}\,$\lambda$4471/\mbox{He\,{\sc ii}}\,$\lambda$4541 is
somewhat steeper for LS~5039, which suggests a slightly earlier spectral type
than O7\,V, in good agreement with the O6.5\,V spectral type proposed by
\cite{clark01}. Therefore, a synthetic O6.5\,V template was generated for the
cross-correlation analysis. This was computed using the NLTE library OSTAR2002
\citep{lanz03} for $T_{\rm eff}=39\,000$~K, $\log~g=3.85$, $v \sin
i=113$~km~s$^{-1}$ (see Sects.~\ref{rotational} and \ref{stellar}) and
degraded to our instrument resolution of 83~km~s$^{-1}$. Every R900V spectrum
of LS~5039 was then cross-correlated with the synthetic spectrum, after
subtracting the continuum. The main IS absorption lines and bands at
$\lambda$3934 (\mbox{Ca\,{\sc ii}} K), $\lambda$4430, $\lambda$4501,
$\lambda$4726, $\lambda$4762, $\lambda$4885, $\lambda$5449 and
$\lambda\lambda$5487--5550 were masked, leaving all Balmer and He lines for
the analysis. We present the resulting velocities in Fig.~\ref{figrvn}. Note
the night-to-night variability which supports a $\approx$4-day periodicity, in
line with the results of M04. Some nights also show evidence for superimposed
short time-scale ($\sim$hr) variability that is likely produced by a
contaminating wind component.

\begin{figure}
\centering
\includegraphics[angle=0,width=84mm]{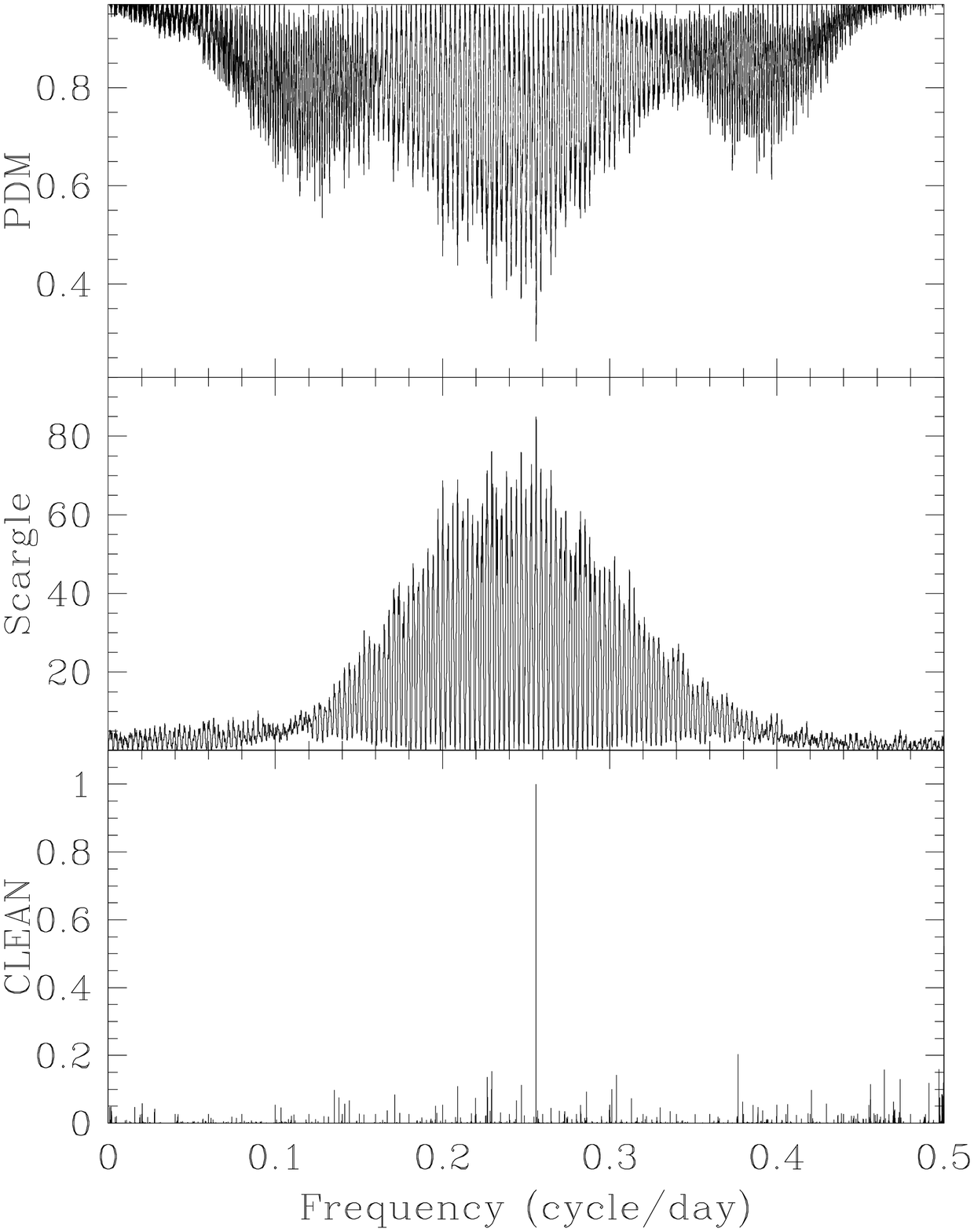}
\caption{Period search results obtained after combining our INT radial
velocities with the ones from M01 and M04. Top: PDM spectrum showing the
deepest minimum at 0.256 cycle~d$^{-1}$. Middle: Scargle power spectrum with
the strongest peak also at 0.256 cycle~d$^{-1}$. Bottom: Periodogram obtained
with the CLEAN algorithm, showing compatible results.}
\label{figper}
\end{figure}

\begin{table*}
\centering
\caption[]{Orbital solutions from an eccentric fit to different data sets. The
All solution stands for a fit performed on our INT plus M01 and M04 data for
all lines. The Balmer, He {\sc i}, and He {\sc ii} solutions have been
obtained with a fixed $P_{\rm orb}$ and $T_0$ on our INT data alone.}
\begin{tabular}{lcccc}
\hline 
Parameter                 & All                   & Balmer                & \mbox{He\,{\sc i}}    & \mbox{He\,{\sc ii}} (adopted)\\
\hline                                           
$P_{\rm orb}$ (days)      & 3.90603 $\pm$ 0.00017 & 3.90603 (fixed)       & 3.90603 (fixed)       & 3.90603 (fixed)      \\
$T_0$ (HJD$-$2\,451\,000) & 943.09  $\pm$ 0.10    & 943.09 (fixed)        & 943.09 (fixed)        & 943.09 (fixed)       \\
$e$                       & 0.31    $\pm$ 0.04    & 0.34    $\pm$ 0.04    & 0.24    $\pm$ 0.06    & 0.35    $\pm$ 0.04   \\
$w$ (\degr)               & 226     $\pm$ 8       & 226.7   $\pm$ 3.3     & 228.0   $\pm$ 4.2     & 225.8   $\pm$ 3.3    \\
$\gamma$ (km~s$^{-1}$)    & 8.1     $\pm$ 0.5     & 9.2     $\pm$ 0.8     & 12.9    $\pm$ 0.8     & 17.2    $\pm$ 0.7    \\
$K_1$ (km~s$^{-1}$)       & 19.4    $\pm$ 0.9     & 30.4    $\pm$ 1.6     & 21.1    $\pm$ 1.3     & 25.2    $\pm$ 1.4    \\
$a_1 \sin i$ (R$_\odot$)  & 1.42    $\pm$ 0.07    & 2.20    $\pm$ 0.12    & 1.58    $\pm$ 0.10    & 1.82    $\pm$ 0.10   \\
$f(M)$ (M$_\odot$)        & 0.0025  $\pm$ 0.0004  & 0.0094  $\pm$ 0.0015  & 0.0035  $\pm$ 0.0007  & 0.0053  $\pm$ 0.0009 \\
rms of fit (km~s$^{-1}$)  & 6.8                   & 11.2                  & 11.5                  & 9.1                  \\
\hline
\label{esolutions}
\end{tabular}
\end{table*}

\begin{figure}
\centering
\includegraphics[angle=0,width=84mm]{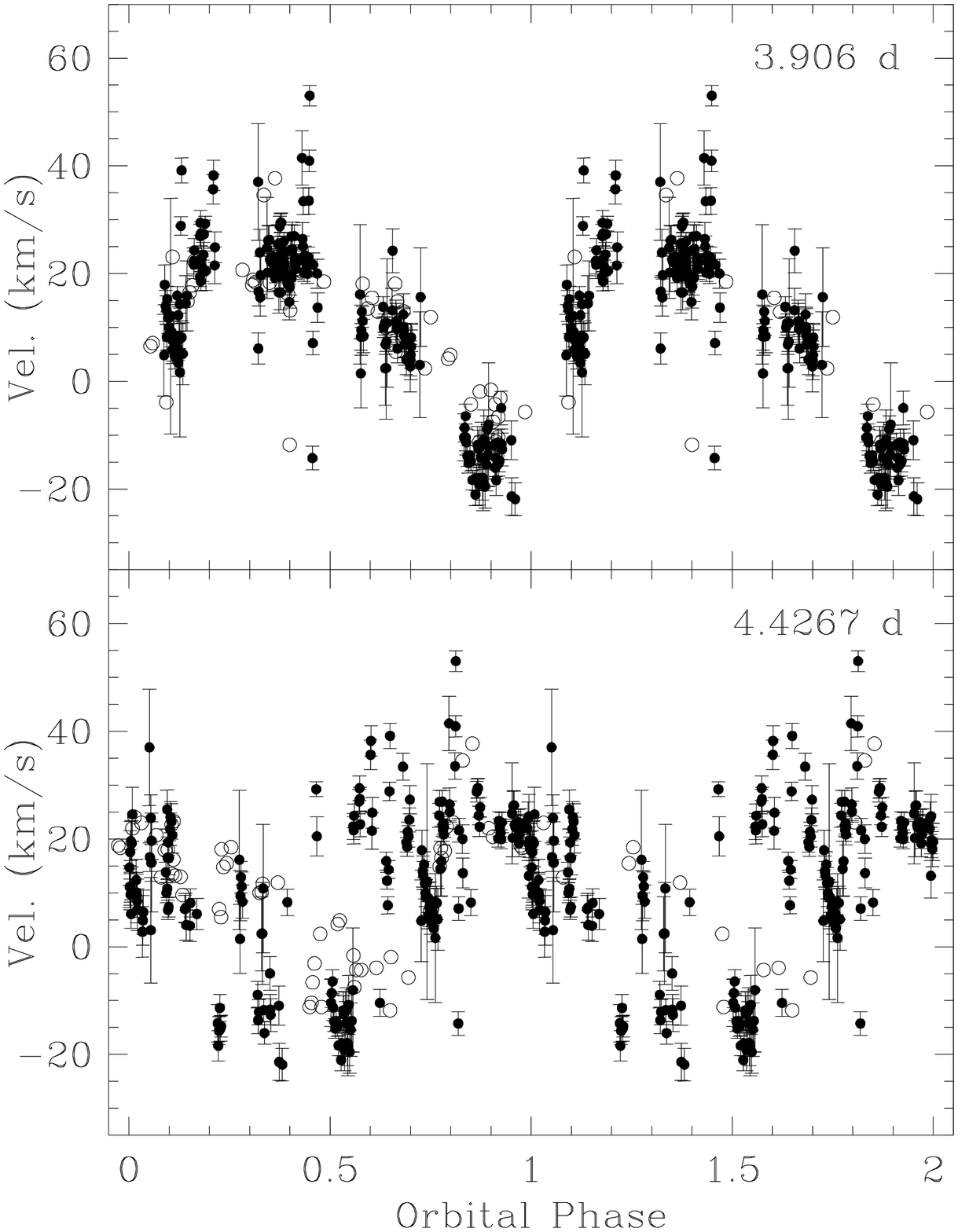}
\caption{INT (filled circles) and McSwain's (open circles) velocities folded
on the 3.906~d and 4.4267~d periods using $T_0$=HJD~2\,451\,943.09. Two
orbital cycles are shown for better display. As can be seen, the new data rule
out the M04 period.}
\label{figfold}
\end{figure}

In order to obtain the orbital period of LS~5039 we have performed a timing
analysis on a database including our radial velocities (196 data points) and
the ones reported in M01 and M04 (54 data points), to get advantage of the
long baseline and to break the 1-year alias of our window function. We have
noted a small systematic offset of $-$6~km~s$^{-1}$ between McSwain's
velocities and ours (obtained by computing the running mean of the two
databases) and this has been corrected before the analysis. Because the
orbital modulation may not be sinusoidal due to a possible eccentricity (see
M01 and M04), we have employed the phase dispersion minimization (PDM)
algorithm (\citealt{stellingwerf78}), that is better suited for non-sinusoidal
periodicities (see, e.g., \citealt{otazu02}, \citeyear{otazu04}). For
completeness we have also used the standard Scargle Fourier Transform
(\citealt{scargle82}) and the CLEAN algorithm (\citealt*{roberts87}). All the
periodograms were computed in the frequency range $\nu = 0.05$--$3$
cycle~d$^{-1}$ with resolution $1 \times 10^{-5}$ cycle~d$^{-1}$. We show in
Fig.~\ref{figper} the power spectra obtained by the three methods and they all
provide significant peaks at a frequency of 0.2560$\pm$0.0001 cycle~d$^{-1}$,
corresponding to 3.906$\pm$0.001~d. We plot in Fig.~\ref{figfold} all the
radial velocities folded on the 4.4267~d period of M04 and our favoured
3.906~d period. We note that although our radial velocity measurements taken
on July 2002 were compatible with the ephemeris of M04, as it is stated in
their paper, the new measurements acquired on July 2003 are not, therefore
ruling out the 4.4267~d period. There are several INT data points that deviate
from the general trend of the radial velocity curve with the 3.906~d period
around phase 0.45, which correspond to the excursion around HJD~2\,452\,827.5
visible in Fig.~\ref{figrvn}. The spurious data point of McSwain around phase
0.4 corresponds to the first value reported in M04, which was obtained from a
single spectrum in that night and could be the result of a short-term
excursion as the one quoted above. Since the radial velocity curve is not
sinusoidal, we subsequently decided to fit an eccentric orbit model (see
\citealt{wilson71} and \citealt{vanhamme03}\footnote{The original Wilson \&
Devinney code has suffered major upgrades since its first release, including
significant improvements in the underlying physical models. The most recent
version of the code together with the relevant documentation can be found in
{\tt ftp://astro.ufl.edu/pub/wilson/lcdc2003}.}) to our database. As McSwain's
velocities have no error bars we assigned equal weight to all the velocities
in the fit. Our best solution yields $P_{\rm orb}$ = 3.90603$\pm$0.00017~d and
$T_0$=HJD~2\,451\,943.09$\pm$0.10 (i.e. time of periastron passage). The rest
of the orbital elements are listed in the first column of
Table~\ref{esolutions}.

\begin{figure}
\centering
\includegraphics[angle=0,width=84mm]{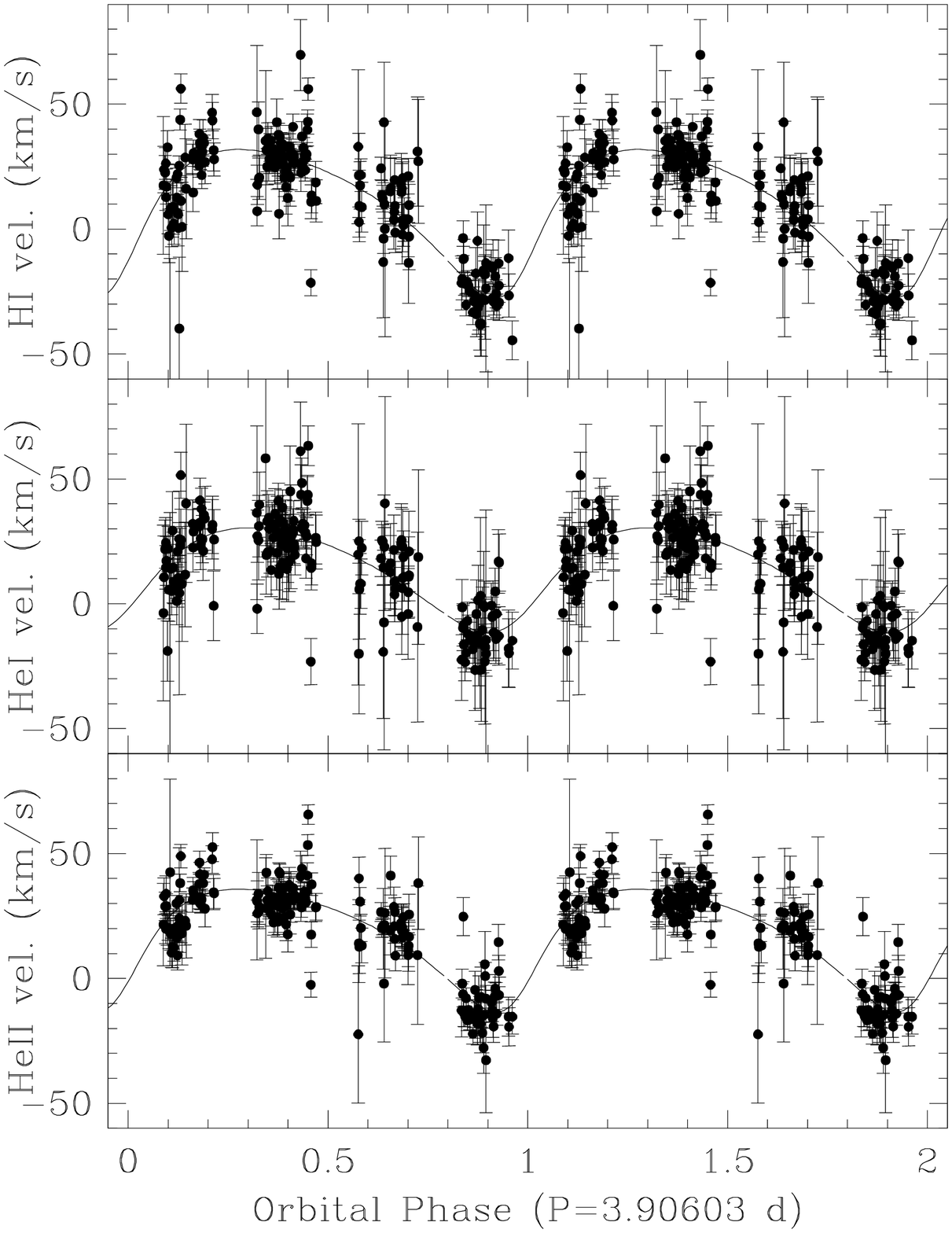}
\caption{INT velocities folded on the 3.90603~d period with
$T_0$=HJD~2\,451\,943.09. The radial velocities have been obtained through
cross-correlating all Balmer (top), \mbox{He\,{\sc i}} (middle) and
\mbox{He\,{\sc ii}} lines (bottom). We superimpose the best eccentric orbital
fits, the parameters of which are quoted in Table~\ref{esolutions}.}
\label{figrv}
\end{figure}

\begin{figure}
\centering
\includegraphics[angle=0,width=84mm]{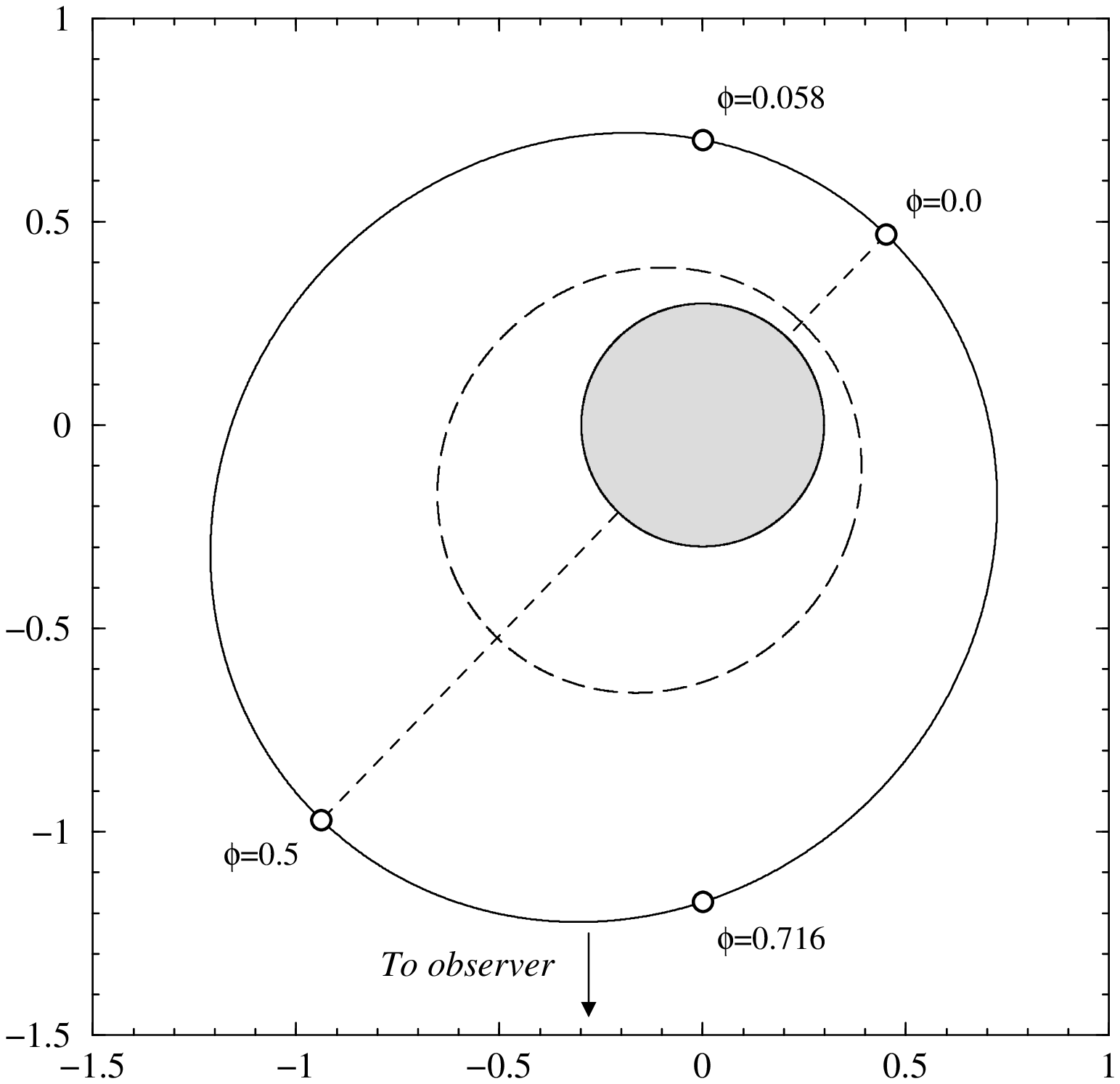}
\caption{Relative orbit as seen from above of the compact object around the
optical component, which lies in the ellipse focus, computed using our adopted
solution from Table~\ref{esolutions}. The coordinates are in units of the
orbital semi-major axis. Relevant phases such as the periastron, apastron, and
conjunctions are indicated, with the dashed line joining periastron and
apastron. The grey circle represents the optical companion to scale in the
case of $R_{\rm O}=9.3$~R$_\odot$, while the long-dashed ellipse represents
the radius of the Roche lobe at each orbital phase for the adopted masses of
$M_{\rm O}=22.9$~M$_\odot$ and $M_{\rm X}=3.7$~M$_\odot$ (see
Sect.~\ref{mass}). The star is at 85 per cent of filling the Roche lobe during
the periastron passage.}
\label{figorbit}
\end{figure}

Since hydrogen lines in early-type stars may be contaminated by wind emission
\citep{puls96}, we decided to treat different line species separately.
Therefore, we extracted sets of radial velocities for the \mbox{H\,{\sc i}},
\mbox{He\,{\sc i}} and \mbox{He\,{\sc ii}} lines from our INT spectra.
Figure~\ref{figrv} presents the radial velocity curves for the three groups of
lines folded on the 3.90603~d period and $T_0$=HJD~2\,451\,943.09. We have
also fitted these velocity curves with eccentric orbital models, but using the
data errors for the weighting scheme and fixing $P_{\rm orb}$ and $T_0$ to the
above values. Table~\ref{esolutions} lists the best-fitting parameters for the
separated groups of lines, and the corresponding orbital solutions can be seen
overplotted in Fig.~\ref{figrv}. We note a trend in the $\gamma$-velocities
towards redder values as one moves from Balmer to the higher excitation
\mbox{He\,{\sc i}} and \mbox{He\,{\sc ii}} lines. Blue-shifted velocities are
usually considered a signature of stellar wind, due to P-Cygni contamination
of the photospheric profiles, and can be conspicuous in the low members of the
Balmer series and some \mbox{He\,{\sc i}} lines, such as $\lambda$4471.
Furthermore, the fitted solutions to the Balmer and \mbox{He\,{\sc i}} lines
present significantly larger rms scatter than the \mbox{He\,{\sc ii}}
solution. This is also consistent with a tentative scenario of variable wind
contamination in Balmer and \mbox{He\,{\sc i}} lines. Therefore, we decided to
give more credit to the \mbox{He\,{\sc ii}} solution, which will be adopted as
the true orbital elements of LS~5039 for the remainder of the paper. As a
test, we have also computed a combined solution using the weighted mean of
every orbital parameter and we find this to be fully consistent with the
\mbox{He\,{\sc ii}} solution within errors, except for the systemic velocity
which is obviously on the blue side. Our $K$-velocity yields a larger $f(M)$
and compact object's mass range than in previous works (M01, M04). To better
illustrate the resulting orbit, we show in Fig.~\ref{figorbit} the relative
motion of the compact object around the optical companion as seen from above
(i.e., for an observer with $i=0\degr$).

\subsection{Rotational Broadening} \label{rotational}

As a first approach, we have followed the technique applied to 1A~0620$-$00 by
\cite*{marsh94} and described in their paper. Essentially, we subtract
different broadened versions of the O7\,V((f)) template HD~168075~B from our
Doppler corrected average spectrum of LS~5039 and perform a $\chi^2$ test on
the residuals. The O7\,V((f)) spectrum was broadened by convolution with the
rotational profile of \cite{gray92} which assumes a linearized limb darkening
coefficient $\epsilon$. We have taken $\epsilon=0.23$ which is appropriate for
the stellar parameters of our star ($T_{\rm eff}=39\,000$~K, $\log~g=3.85$,
see Sect.~\ref{stellar}) in the $B$-band. We performed the analysis
independently for the two groups of spectra at different resolutions, 83 and
30 km~s$^{-1}$, and find minimum $\chi^2$ for broadenings of $105\pm2$ and
$126\pm12$~km~s$^{-1}$, respectively. Since the template also has a rotational
velocity of 79~km~s$^{-1}$ (\citealt{penny96}), we need to sum this
quadratically to get the true rotational velocity of LS~5039. This yields $v
\sin i=131\pm2$ and $149\pm12$ km~s$^{-1}$ for the two resolutions. These
numbers can be compared with previous determinations of $131\pm6$~km~s$^{-1}$
(M01) and $140\pm8$~km~s$^{-1}$ (M04).

Alternatively, we attempted a refined determination following the Fourier
technique described by \cite{gray92}; for a recent application see
\cite{royer02}. This technique takes advantage of the fact that the Fourier
transform of the rotational profile has zeroes at regular positions that
depend on the projected rotational velocity. The whole line profile, being a
convolution of different contributing profiles in the wavelength domain,
conserves these zeroes in the Fourier domain. While the natural profile does
not introduce additional zeroes, other broadening mechanisms may do, although
at the large rotational velocities of the O stars the firsts zeroes of the
transformed profile are expected to be due to rotation. However, as the Stark
effect strongly dominates the line profiles of Balmer and \mbox{He\,{\sc ii}}
lines, we have only used the \mbox{He\,{\sc i}} lines for the $v \sin i$
determination. This method yields $v \sin i = 113\pm8$~km~s$^{-1}$, with the
error obtained from the dispersion of the individual lines. This is a
significantly lower projected rotational velocity than any of the previously
obtained ones. However, this lower value is consistent with the finding that a
part of the line broadening usually attributed to rotation is probably due to
some kind of turbulence (see for example \citealt{ryans02}) and, therefore, we
give more credit to this latter determination. We note that for the fit to the
line profiles (next section) we have used a projected rotational velocity of
150~km~s$^{-1}$ in good agreement with M04.

\subsection{Stellar Parameters} \label{stellar}

We have determined the stellar parameters using the latest version of FASTWIND
(\citealt*{santolaya97}; \citealt*{repolust04}; \citealt{puls05}). This is a
NLTE, spherical, mass-losing model atmosphere code particularly optimized for
the analysis of massive OBA stars. We have fitted simultaneously the following
lines: H$\alpha$, H$\beta$, H$\gamma$, H$\delta$, \mbox{He\,{\sc
i}}~$\lambda$4387, $\lambda$4471, $\lambda$4922, and \mbox{He\,{\sc
ii}}~$\lambda$4200, $\lambda$4541, $\lambda$4686. From the analysis we obtain
$T_{\rm eff}=39\,000\pm1\,000$~K and $\log~g=3.85\pm0.10$. While our
temperature is consistent with that of M04 our gravity is slightly lower than
the one adopted by these authors (although adopted errors allow for a marginal
agreement). Note, however, that their gravity, which was determined by model
fitting only to the wings of the H$\gamma$ profile, is likely biased upward by
the blend of \mbox{N\,{\sc iii}} lines (see details in M04). We also confirm
that LS~5039 is contaminated by CNO products (see M04), with strong N and weak
C lines, as compared to stars of similar spectral type. It can therefore be
classified as a member of the ON type group stars, i.e., its complete spectral
type is ON6.5\,V((f)), as given by M04.

Adopting the best-fitting extinction parameters given by M04 (i.e.
$E(B-V)=1.28\pm0.02$ and $R=3.18\pm0.07$ which leads to $A_{\rm
V}=4.07\pm0.11$, different from what they quote), the updated $M_{\rm
V}=-4.77\pm0.15$ for an O6.5\,V star from the new calibration by
\cite*{martins05}, and $V=11.33\pm0.02$ and $V=11.32\pm0.01$~mag from
\cite{clark01} we derive a distance of $2.54\pm0.04$~kpc. We have also
considered the slightly different values of $V$ reported in the literature,
ranging from 11.20 to 11.39~mag with typical 1$\sigma$ uncertainties of
0.03~mag (\citealt{drilling91}; \citealt{lahulla92}; \citealt{marti98};
\citealt{marti04}), and computed the corresponding distances. The weighted
mean and standard deviation of all these values provides a more realistic
distance estimate of $d=2.5\pm0.1$~kpc, that will be adopted for the remainder
of the paper. With these numbers we get a radius of $R_{\rm
O}=9.3^{+0.7}_{-0.6}$~R$_\odot$ (see, e.g. \citealt*{herrero02}). From this
radius we derive a luminosity of $\log (L_{\rm O}/$L$_\odot)=5.26\pm0.06$ and
a mass of $M_{\rm O}=22.9^{+3.4}_{-2.9}$~M$_\odot$.

The mass-loss rate has been derived from the H$\alpha$ profile, which appears
to be variable on a secular time-scale (\citealt{reig03}; M04;
\citealt{bosch05}). Our 2003 data show a very stable equivalent width with a
mean value of $EW=2.8\pm0.1$~\AA\ and no variations above a 5 per cent level
over 1 orbital cycle. Despite this, we have selected two extreme profiles
showing minimum and maximum $EW$ for the calculation, that we have performed
by using again the latest version of FASTWIND (\citealt{puls05}; see examples
of use in \citealt{repolust04}). For the lower state (i.e., largest
absorption) we obtain a wind mass-loss rate of $3.7\times10^{-7}$
M$_\odot$~yr$^{-1}$. While we can put an upper limit for this state of
$5.0\times10^{-7}$ M$_\odot$~yr$^{-1}$, the lower limit is much more
uncertain, as the sensitivity of H$\alpha$ to mass-loss rate falls off rapidly
at such low values. For the high state, we obtain a best value of
$7.5\times10^{-7}$ M$_\odot$~yr$^{-1}$, i.e., a factor of two if we interpret
the small difference in the profiles as due to the stellar wind and not to the
low signal-to-noise ratio. Again, $5.0\times10^{-7}$ M$_\odot$~yr$^{-1}$ can
be considered a lower limit for this state, while we can set an absolute upper
limit to the high state mass-loss rate of $1.0\times10^{-6}$
M$_\odot$~yr$^{-1}$. With these values, the Modified Wind Momentum of the
upper state coincides with the values quoted by \cite{repolust04}, being a
factor of two lower for the low state, but still within their uncertainties.
We also note that our values are a factor $\sim$4--8 higher than those derived
by M04 from data acquired in 2002, but compare well with average mass-loss
rates in O6.5\,V stars \citep{howarth89}.

\section{Mass of the compact object} \label{mass}

\begin{figure}
\centering
\includegraphics[angle=0,width=84mm]{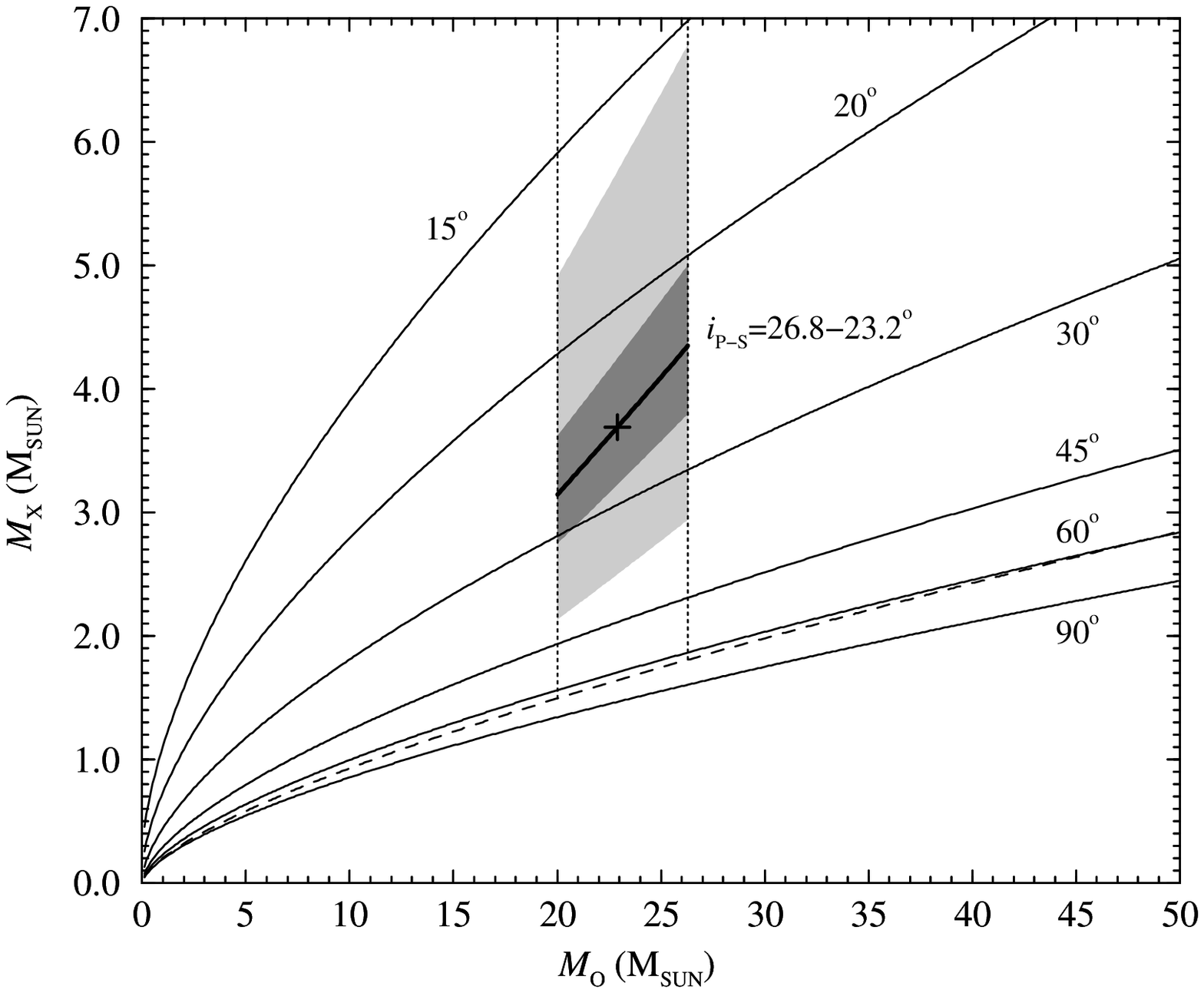}
\caption{Mass constraints for the two stars in LS~5039 derived from our
orbital solution. The thin solid lines have been calculated by using our
adopted mass function, $f(M)=0.0053$~M$_\odot$, and the quoted inclination
angles. The dashed line indicates the lower limit of $M_{\rm X}$ from the
absence of X-ray eclipses. The dotted lines enclose the area of valid
solutions for the interval of likely values of $M_{\rm O}$. The thicker line
represents the valid solutions by assuming pseudo-synchronization, while the
dark-grey region represents all their possible 1$\sigma$ errors, and the
light-grey region their 3$\sigma$ errors. The cross indicates the case of
$M_{\rm O}=22.9$~M$_\odot$, which implies $i_{\rm P-S}=24.9\degr$ and $M_{\rm
X}=3.7$~M$_\odot$.}
\label{figmasses}
\end{figure}

Our revised mass function and stellar parameters provide new constraints on
the mass of the compact object in LS~5039. First of all, considering the
adopted value of $f(M)=0.0053\pm0.0009$~M$_\odot$ and using an inclination of
$i=90\degr$ we obtain a strict lower limit for the mass of the compact object
as a function of $M_{\rm O}$. This limit is represented by the lowest solid
line in Fig.~\ref{figmasses}, where we also plot similar relationships for
other inclination angles. For the range of possible masses of the optical
companion, $22.9^{+3.4}_{-2.9}$~M$_\odot$, we obtain a lower limit on the mass
of the compact object in the range 1.34--1.61~M$_\odot$, with an additional
linear 1$\sigma$ uncertainty of 0.11~M$_\odot$ due to the mass function error
bar.

Another constraint comes from the fact that no X-ray eclipses are seen in
LS~5039. With the new ephemeris reported here, the {\it BeppoSAX} observations
performed by \cite{reig03} cover phases between 0.969 and 0.205, while an
eventual X-ray eclipse should be centered at phase 0.058 (see
Fig.~\ref{figorbit}). In fact, periastron takes place at
$t=(1.0\pm0.9)\times10^4$~s in figure~1 of \cite{reig03}, while superior
conjunction of the compact object takes place at $t=(3.0\pm0.9)\times10^4$~s,
where the errors come from the uncertainty in our $T_0$. The X-ray flux is
nearly constant during this last interval, implying that X-ray eclipses can be
definitively ruled out in LS~5039. This condition has been used to compute the
lower limit on $M_{\rm X}$ as a function of $M_{\rm O}$ (which in turn is also
a function of $R_{\rm O}$). This is represented as a dashed line in
Fig.~\ref{figmasses}, which yields $i<64.6\degr$ and $M_{\rm
X}>1.49$~M$_\odot$ for the case of $M_{\rm O}=20.0$~M$_\odot$, and
$i<63.3\degr$ and $M_{\rm X}>1.81$~M$_\odot$ for $M_{\rm O}=26.3$~M$_\odot$.
Therefore, this condition provides a compact object mass above
$1.49\pm0.11$~M$_\odot$ (the errorbar due to the $f(M)$ uncertainty). This
value is still below the 1.75--2.44~M$_{\odot}$ neutron star mass in Vela~X-1
\citep{quaintrell03}, or the (95 per cent confidence) lower limit of
1.68~M$_{\odot}$ for at least one of the two pulsars in binaries with
eccentric orbits within the globular cluster Terzan~5 \citep{ransom05}.
Therefore, all the previously obtained values are compatible with a massive or
even a canonical neutron star.

A further constraint can be obtained by assuming that the optical companion
star is pseudo-synchronized, i.e. its rotational and orbital angular
velocities are synchronized at periastron passage (\citealt{kopal78};
\citealt{claret93}). We have estimated the synchronization time-scale of the
optical component of LS~5039 using the formalism by \cite{zahn89} (as
formulated in \citealt{claret97}) and the tidal evolution parameters of the
stellar models of \cite{claret04}. We find a synchronization time-scale of
about 1~Myr. This value is similar to the (rough) upper limit to the age of 
LS~5039 \citep{ribo02} and, therefore, the system appears to have had time to
reach orbital pseudo-synchronism. Note, however, that both the synchronization
timescale and the age are only accurate as an order of magnitude estimate and,
consequently, we can only state that both quantities do not exclude each
other. On the other hand, the theoretical orbital circularization time-scale
comes out to be a much larger value of 10~Myr, compatible with the observation
of an eccentric orbit. Therefore, assuming pseudo-synchronization (P-S) we can
combine our determination of $v \sin i$ and radius to estimate the binary
inclination $i_{\rm P-S}$, i.e. \[ v \sin i = 2 \pi~F~R_{\rm O}~ \sin i_{\rm
P-S}~ P_{\rm orb}^{-1}, \] where $F=(1+e)^{1/2}/(1-e)^{3/2}=2.22\pm0.17$, and
we find values of $i_{\rm P-S}$ in the range 26.8--23.2\degr, depending on
$M_{\rm O}$. These inclinations set strong constraints to the compact object's
mass, as is displayed by the thicker solid line in Fig.~\ref{figmasses}. The
possible values of $M_{\rm X}$ range from 3.14 to 4.35~M$_{\odot}$. The
1$\sigma$ uncertainty region (obtained through propagating errors in $i_{\rm
P-S}$ and $f(M)$) is marked by the dark-grey area in Fig.~\ref{figmasses}.
This leads to compact object masses in the range 2.75--5.00~M$_{\odot}$. The
3$\sigma$ uncertainty region is also indicated in Fig.~\ref{figmasses} with a
light-grey area and yield masses in the range 2.14--6.78~M$_{\odot}$. The
central value of $M_{\rm O}=22.9$~M$_{\odot}$ implies $i_{\rm
P-S}=24.9\pm2.8\degr$ (1$\sigma$ uncertainty) and $M_{\rm
X}=3.7^{+1.3}_{-1.0}$~M$_{\odot}$ (1$\sigma$ uncertainty in all involved
parameters), and is indicated by a cross in Fig.~\ref{figmasses}. Therefore,
if the plausible assumption of pseudo-synchronization is correct, the compact
object is consistent with being a black hole.

Since there is no significant H$\alpha$ emission and the binary system does
not exhibit strong X-ray outbursts during the periastron passage (see, e.g.,
\citealt{bosch05}), the donor star is not expected to overflow its Roche lobe
at any orbital phase. To check this, we have generated a grid of solutions in
the following way: for each possible optical companion mass in
Fig.~\ref{figmasses} we have iterated for all possible inclination angles and
obtained the corresponding $M_{\rm X}$ (and its error) through the mass
function. Then $M_{\rm X}$ is used to compute the Roche lobe at periastron 
and its uncertainty (using the semi-analytical expression in 
\citealt{eggleton83}). We find that, for the possible range of optical star 
masses $M_{\rm O}=20.0$--$26.3$~M$_{\odot}$ and within 1$\sigma$, the Roche
lobe is not overflown at periastron for $M_{\rm X}<8$~M$_{\odot}$ (or
$i>13.4\degr$).

A final constraint on the inclination angle comes from the rotational 
velocity. The break-up speed, or critical rotational velocity, can be obtained
by means of $v_{\rm crit}=\sqrt{(2/3)GM_{\rm O}/R_{\rm O}}$ (see
\citealt{porter96}). Considering the limits of our likely values for the mass
and radius of the optical companion we obtain $v_{\rm
crit}=540$--580~km~s$^{-1}$. Finally, by using our measured rotational
velocity of $v \sin i = 113\pm8$~km~s$^{-1}$, we can constrain the inclination
angle to be above 11--12\degr, which in turn provides upper limits for the
compact object mass of 8--10~M$_{\odot}$ (depending on $M_{\rm O}$).

In summary, the lack of X-ray eclipses, Roche lobe overflow, and break-up
speed constrain the inclination angle to be in the range 13--64\degr, and the
compact object mass to 1.5--8~M$_{\odot}$. If the primary in LS~5039 is
pseudo-synchronized the inclination angle is $24.9\pm2.8\degr$, and the
compact object mass $M_{\rm X}=3.7^{+1.3}_{-1.0}$~M$_{\odot}$.

The inclination angle could be tested through accurate photometry near the
periastron passage. We have simulated the light curve of the tidally distorted
optical component using our orbital parameters and predict a $\sim$0.02~mag
increase in brightness near the periastron for $i=45\degr$, which could be
revealed by 2--3 mmag accuracy photometry. If this variability is not detected
or is below 0.01~mag, the inclination angle must be below 30\degr, and the
mass of the compact object above 3.0~M$_{\odot}$, confirming its black hole
nature. We have looked for this effect in our existing photometry
\citep{marti04} but the results are not conclusive due to noise and scarce
phase sampling. More observations are currently underway.

\section{Discussion} \label{discussion}

In this section we evaluate the implications of all the updated parameters on
several properties of LS~5039.

\subsection{Optical companion} \label{companion}

The contamination of the optical companion by CNO products may be a
consequence of mass transfer of CNO-processed material from the supernova
progenitor prior to the explosion (M04). Alternatively, it might be caused by
mixing processes (induced by rotation) which brings CNO processed material
from the inner regions into the stellar atmosphere \citep*{heger00}. Indeed,
if the system is pseudo-synchronized, then our measured $v \sin i =
113\pm8$~km~s$^{-1}$ translates into a large rotational velocity of
$v=268\pm34$~km~s$^{-1}$, which is $\sim$0.5$v_{\rm crit}$. Therefore,
although we cannot discard the hypothesis of mass transfer prior to the SN
explosion, we favour the mixing scenario to explain the presence of CNO
products. We also note that the optical companion in LS~5039 is slightly
undermassive for its spectral type, since from the calibration of
\cite{martins05} an O6.5\,V star should have $M_{\rm spec}=29.0$~M$_\odot$,
while our measured mass of $M_{\rm O}=22.9^{+3.4}_{-2.9}$~M$_\odot$ covers
spectral types between O7\,V and O8.5\,V in their calibration. This effect has
been observed in other HMXB (see, e.g., \citealt{vdheuvel83}).

\subsection{X-ray variability and accretion/ejection energetic balance} \label{balance}

An interesting consequence of our orbital solution is the smaller
eccentricity, $e=0.35\pm0.04$, compared to McSwain's solution,
$e=0.48\pm0.06$. A lower eccentricity is easier to reconcile with the weak
orbital X-ray modulation, with the X-ray flux varying a factor of 2.5,
observed by \cite{bosch05}. In a simplistic Bondi-Hoyle scenario (i.e.
spherical accretion through winds) our eccentricity, together with the updated
masses, provides an X-ray variability with a factor of 12 between maximum and
minimum flux, which is still a factor $\approx$5 higher than observed (see
\citealt{reig03} for details about the method). Even when considering that all
possible uncertainties conspire in one or the opposite direction, the expected
variability is in the range 7--24, which is a factor 3--10 higher than
observed. This led \cite{bosch05} to propose the existence of an accretion
disc that would partly screen the compact object from direct wind accretion.

We can now compare the luminosity released in the vicinity of the compact 
object (i.e., excluding the contribution from the companion which dominates 
in the optical domain) with the available accretion luminosity. The first one
includes the radio, X-ray, high energy, and very high energy (VHE) gamma-ray
domains. It is maximum in the high energy gamma-ray domain since, using our
new distance estimate of 2.5~kpc, we have: $L_{\rm
radio,~0.1-100~GHz}=8\times10^{30}$ \citep{marti98}, $L_{\rm
X,~3-30~keV}=1.0$--$2.5\times10^{34}$ (\citealt{bosch05}, after removing the
diffuse background contamination), $L_{\gamma,~> 100~{\rm
MeV}}=2.7\times10^{35}$ \citep{hartman99}, and $L_{{\rm VHE},~> 250~{\rm
GeV}}=4\times10^{33}$~erg~s$^{-1}$ \citep{aharonian05}. Regarding the
accretion luminosity, we have used the Bondi-Hoyle scenario to obtain a rough
estimate of the accreted matter and the energy that can be released from the
accretion process. By using the updated parameters ($M_{\rm
O}=22.9$~M$_\odot$, $R_{\rm O}=9.3$~R$_\odot$, $M_{\rm X}=3.7$~M$_\odot$,
$R_{\rm X}=R_{\rm Sch}=2GM_{\rm X}/c^2=2.95\,(M_{\rm X}/{\rm M}_\odot)$~km,
$P_{\rm orb}=3.90603$~d, $e=0.35$,
$\dot{M}=5\times10^{-7}$~M$_\odot$~yr$^{-1}$), $v_{\rm inf}=2440$~km~s$^{-1}$
from M04 and $\beta=0.8$ we obtain an accretion luminosity averaged through
the orbit of $<L_{\rm acc}>=8\times10^{35}$~erg~s$^{-1}$. Therefore, around
1/3 of the accreted luminosity would be radiated within the relativistic jet
of LS~5039, while the remaining available luminosity could be lost by the
advection of matter towards the black hole. Of course, the scenario is not so
simple. First of all, we should include the kinetic luminosity of the jet,
given by $L_{\rm k}=(\Gamma-1)\dot{M}_{\rm jet}c^2$ (where $\Gamma$ is the
bulk Lorentz factor of the jet, and we have neglected the energy needed to
abandon the potential well). For a jet with a velocity $\beta=v/c=0.2$ (from
$\beta\cos\theta=0.17\pm0.05$ found by \citealt{paredes02} and assuming
$\theta=i_{\rm P-S}\simeq25\degr$), or $\Gamma=1.02$, we obtain $L_{\rm
k}=1.0\times10^{36}$~erg~s$^{-1}$ (see details on how to estimate
$\dot{M}_{\rm jet}$ in \citealt{paredes00}). This luminosity is slightly
higher than the accretion luminosity, although this discrepancy could be
solved by increasing the mass of the compact object up to 5~M$_\odot$ and/or
the mass-loss rate of the primary up to $10^{-6}$~M$_\odot$~yr$^{-1}$.
However, the Bondi-Hoyle accretion scenario is an oversimplification of the
real accretion processes that take place in this source, because: 1) the
presence of a thick accretion disc around the compact object seems to be
necessary to launch the relativistic jets in microquasars (see, e.g.
\citealt{fender05} and references therein); 2) the presence of a disc also
appears to be necessary to explain the weak orbital X-ray variability in
LS~5039 \citep{bosch05}. Moreover, we note that due to fast rotation of the
optical star and its proximity to fill the Roche lobe during periastron
passage, we cannot discard additional mass-loss in the equatorial plane of the
binary system, which could provide the needed amount of additional accretion
luminosity. We note that we have neglected the energy stored in magnetic
fields through all this discussion. In any case, we can say that the current
estimates of accretion/ejection luminosities agree quite well by using our new
set of parameters. We note that if the compact object were a neutron star with
$M_{\rm X}=1.4$~M$_\odot$ and $R_{\rm X}=10$~km we would have $<L_{\rm
acc}>=5\times10^{34}$~erg~s$^{-1}$. This value is more than one order of
magnitude lower than for the case of a 3.7~M$_\odot$ black hole, and $\sim$2.5
times smaller than the gamma-ray luminosity of LS~5039.

\subsection{Before and after the SN explosion} \label{supernova}

Using our new radial velocity of the binary system
$\gamma=17.2\pm0.7$~km~s$^{-1}$ and an updated proper motions estimate of
$\mu_{\alpha\cos\delta}=4.8\pm0.8$~mas~yr$^{-1}$,
$\mu_{\delta}=-10.9\pm0.9$~mas~yr$^{-1}$ (including a new radio position
obtained with VLA+Pie Town observations; Mart\'{\i}, Rib\'o \& Paredes, in
preparation), we can recompute the total systemic velocity of LS~5039. We find
$v_{\rm sys}=126\pm9$~km~s$^{-1}$ (see details on the method in
\citealt{ribo02}). On the other hand, our improved masses and eccentricity
have an impact on the formation history of LS~5039. Tidal forces act to
circularize the binary orbit and hence the current eccentricity
$e=0.35\pm0.04$ can be taken as a lower limit to $e_{\rm post-SN}$, the
post-SN eccentricity. In the context of a symmetric SN explosion, $e_{\rm
post-SN}$ is related to the mass lost in the SN event $\Delta M$ through
$\Delta M = e_{\rm post-SN} \times (M_{\rm X} + M_{\rm O})$ which yields
$\Delta M>9\pm2$~M$_{\odot}$ and $P_{\rm re-circ}=3.2\pm0.2$~d (using our
current values for $P_{\rm orb}$ and $e$). With these numbers, and using the
equations in \cite{nelemans99}, we obtain a theoretical recoil velocity of
$130\pm20$~km~s$^{-1}$, in good agreement with our new space velocity. The
previous discrepancy between these two values reported by M04 vanishes thanks,
mainly, to the lower value of $e$, since the equations are only slightly
sensitive to $M_{\rm X}$. Therefore a high mass loss of $\sim$9~M$_{\odot}$
during the SN explosion could provide both the eccentricity and the space
velocity that we currently observe in LS~5039. We note that our predicted
orbital period prior to the SN explosion of 1.8~d yields a Roche lobe radius
of $\sim8.7$~R$_{\odot}$, which is compatible within errors to the O star
radius. Finally, the relatively large mass of the SN progenitor, around
13~M$_{\odot}$, is compatible with the $\la10$--15~M$_{\odot}$ upper limit
found by \cite{fryer01}.

The kinetic energy of the binary system is $4.2\pm0.8\times10^{48}$~erg, i.e.,
merely $4\times10^{-3}$ times the energy of a typical SN, although nearly one
order of magnitude higher than in GRO~J1655$-$40 \citep{mirabel02}. On the
other hand, from the pulsar birth velocity value of $\sim400\pm40$~km~s$^{-1}$
\citep{hobbs05}, and assuming a pulsar mass of 1.4~M$_{\odot}$, we obtain that
their typical linear momentum is $\sim560\pm56$~M$_{\odot}$~km~s$^{-1}$. In
contrast, the linear momentum of the LS~5039 binary system is
$3\,350\pm450$~M$_{\odot}$~km~s$^{-1}$, well above the previous value and
above the value of any individual pulsar. To our knowledge, this value is also
significantly higher than in all other binary systems with measured
velocities, being a possible exception 4U~1700$-$37. For this system, moving
at $70\pm5$~km~s$^{-1}$, and using different masses from \cite{ankay01} and
\cite{clark02}, we obtain a linear momentum in the range
2300--4200~M$_{\odot}$~km~s$^{-1}$, encompassing the value found for LS~5039.

We note that once the orbit is re-circularized, with a period of 3.2~d, the
expected Roche radius will be 15~R$_{\odot}$, well above the current radius of
the star. After leaving the main sequence, the star will fill its Roche lobe,
leading to unstable mass transfer that will probably turn-off the X-ray binary
and microquasar phase of LS~5039 \citep*{frank02}. From the evolutionary 
tracks by \cite{meynet94} we know that the main sequence lifetime is between 
4.5 and 6.5~Myr (assuming a solar metallicity of $Z=0.02$ and for stars with 
initial masses between 40 and 25~M$_{\odot}$). On the other hand, the LS~5039 
trip from the galactic midplane to its current position would take
$\sim$0.5~Myr (explained in \citealt{ribo02} and independent of the distance
to the source). Consequently, the X-ray binary could survive as a microquasar
up to 4--6~Myr from now, placing the system at galactic latitudes of $-10$ to
$-13\degr$. Therefore, relatively nearby systems similar to LS~5039 could be
the counterparts of unidentified EGRET sources with $|b|>5\degr$.

\subsection{Orbital behaviour of the TeV counterpart} \label{hess}

\begin{figure}
\centering
\includegraphics[angle=0,width=84mm]{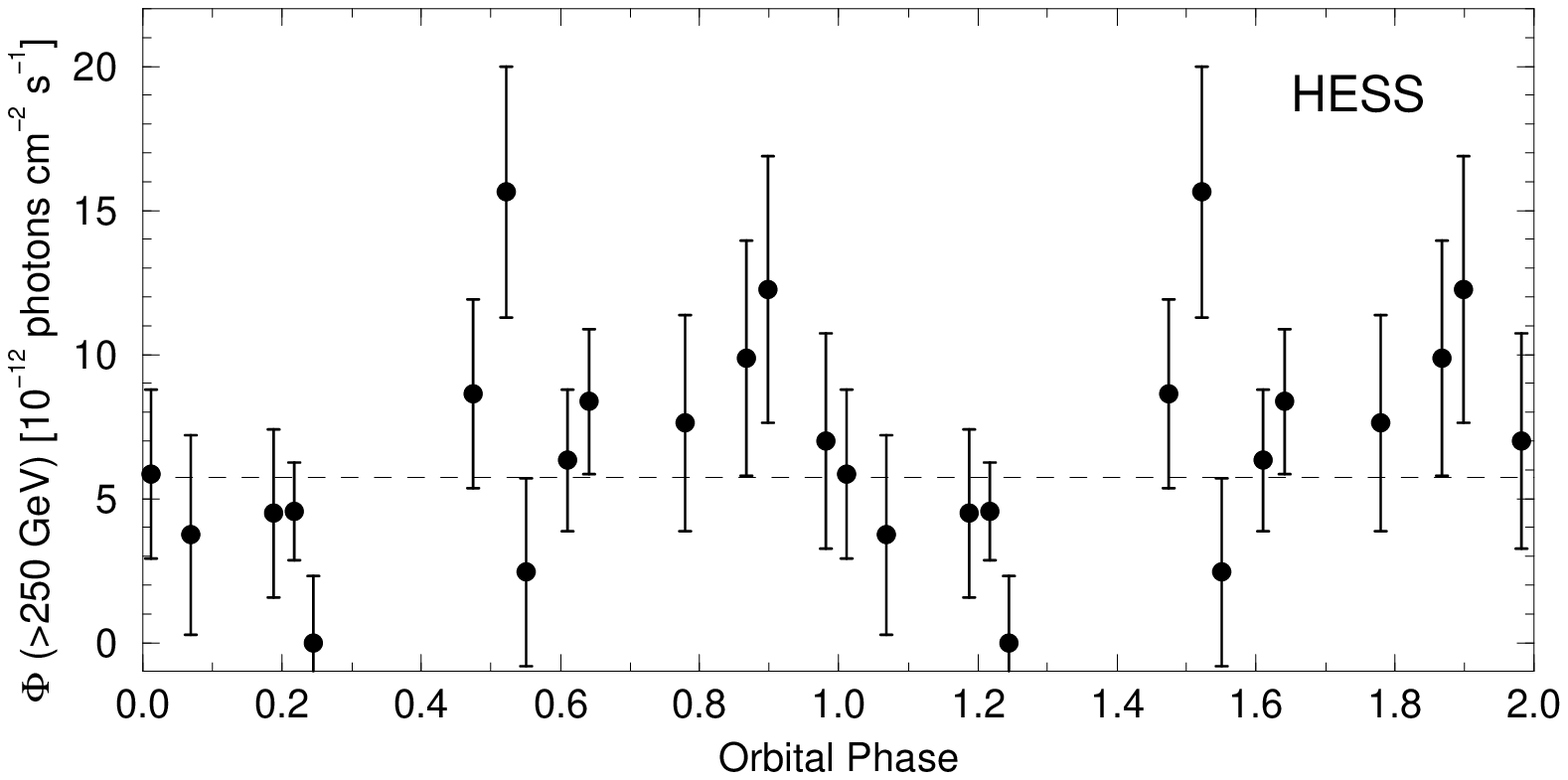}
\caption{VHE gamma-ray fluxes of HESS~J1826$-$148, the proposed TeV
counterpart of LS~5039, folded with our orbital ephemeris ($P_{\rm
orb}=3.90603$~d with $T_0$=HJD~2\,451\,943.09). The dashed line represents the
weighted mean of all data points. Despite the large error bars, there seems to
be a quasi-sinusoidal variation with a maximum around phase 0.9, a similar
behaviour to the one seen in X-rays.}
\label{fighess}
\end{figure}

\cite{aharonian05} have recently reported the detection of a very high energy
gamma-ray source at TeV energies, namely HESS~J1826$-$148, with a position
consistent at the 3$\sigma$ level with that of LS~5039. Moreover, the spectral
energy distribution at high energies makes the association between the HESS
and EGRET sources virtually certain. These authors state that no periodic
variations are apparent when folding the data using the orbital ephemeris of
M04. We show in Fig.~\ref{fighess} the HESS data folded using our new orbital
period. Despite the large error bars, we notice a possible flux variation of a
factor $\sim$3 with the orbital period, with a quasi-sinusoidal pattern and
maximum around phase 0.9. This behaviour is reminiscent to the one recently
found in X-rays in RX~J1826.2$-$1450/LS~5039 by \cite{bosch05}, with a flux
variation of a factor $\sim$2.5 and a maximum around phase 0.8. Although
further HESS observations would be necessary to confirm this orbital
variability, this similarity reinforces the association between LS~5039 and
HESS~J1826$-$148 and, therefore, practically confirms the association between
the microquasar and the EGRET source.

\section{Summary} \label{summary}

We have reported new optical spectroscopy of the microquasar LS~5039 and
obtained a new orbital solution. In particular, we find $P_{\rm
orb}=3.90603\pm0.00017$~d, $e=0.35\pm0.04$, systemic velocity
$\gamma=17.2\pm0.7$~km~s$^{-1}$ and a mass function for the compact object
$f(M)=0.0053 \pm 0.0009$~M$_\odot$, significantly different from previous
results. We have also derived a new distance estimate of $d=2.5\pm0.1$~kpc and
a mass of the optical companion of $M_{\rm O}=22.9^{+3.4}_{-2.9}$~M$_\odot$.
Using this information and assuming pseudo-synchronization we obtain an
inclination of $i=25\pm3\degr$, which yields to $M_{\rm
X}=3.7^{+1.3}_{-1.0}$~M$_{\odot}$. This strongly suggests that the compact
object in LS~5039 is a black hole. With our new orbital parameters
there is a good agreement between the accretion and ejection luminosities
around the compact object. The space velocity of the binary system is also in
good agreement with the theoretical recoil velocity in a symmetric SN
explosion with a mass loss of $\sim$9~M$_{\odot}$. Finally, the orbital
variability of the TeV counterpart is reminiscent to the one seen in X-rays,
reinforcing the association between LS~5039, HESS~J1826$-$148 and
3EG~J1824$-$1514.

\section*{Acknowledgments}

We acknowledge I.~Negueruela for help in the observations and obtaining some 
spectra in the 2002 and 2003 campaigns. We also thank V. Bosch-Ramon,
E.~K\"ording, and L.~J. Pellizza for useful discussions. J.~C. acknowledges
support from the Spanish MCYT grant AYA2002-0036. M.~R., J.~M.~P. and J.~M.
acknowledge partial support by DGI of the Spanish Ministerio de Educaci\'on y
Ciencia (former Ministerio de Ciencia y Tecnolog\'{\i}a) under grants
AYA2001-3092, AYA2004-07171-C02-01 and AYA2004-07171-C02-02, as well as
additional support from the European Regional Development Fund (ERDF/FEDER).
M.~R. acknowledges support from the French Space Agency (CNES) and by a Marie
Curie Fellowship of the European Community programme Improving Human Potential
under contract number HPMF-CT-2002-02053. I.~R. acknowledges support from the
Spanish Ministerio de Ciencia y Tecnolog\'{\i}a through a Ram\'on y Cajal
fellowship. J.~M. is also supported by the Junta de Andaluc\'{\i}a (Spain)
under project FQM322. A.~H. acknowledges support from the Spanish Ministerio
de Educaci\'on y Ciencia grant AYA2004-08271-02-01. MOLLY and DOPPLER software
developed by T.~R. Marsh is gratefully acknowledged. The INT is operated on
the island of La Palma by the Royal Greenwich Observatory in the Spanish
Observatorio del Roque de Los Muchachos of the Instituto de Astrof\'{\i}sica
de Canarias. This research has made use of the NASA's Astrophysics Data System
Abstract Service and of the SIMBAD database, operated at CDS, Strasbourg,
France.

\bsp

\label{lastpage}

\end{document}